\documentclass[aps,prl,reprint,twocolumn,superscriptaddress]{revtex4-2}
\usepackage[english]{babel}

\usepackage{amsmath} % include math package
\usepackage{amssymb} 
\usepackage{graphicx}% Include figure files
\usepackage{dcolumn}% Align table columns on decimal point
\usepackage{bm}% bold math
\usepackage[]{hyperref}
\hypersetup{
    colorlinks=true,
    linkcolor=blue,
    filecolor=magenta,      
    urlcolor=cyan,
    citecolor = blue
    }

\usepackage{color} % for colored text
\usepackage{siunitx}
\usepackage{comment}
\usepackage{cleveref}  % removed [capilatise] option
\crefname{section}{Sec.}{Secs.}%
\usepackage{soul,xcolor}
\usepackage[normalem]{ulem}

\usepackage[nearskip=0pt,farskip=0pt,captionskip=0pt,caption=false]{subfig}
\crefrangelabelformat{figure}{#3#1#4--#5(\crefstripprefix{#1}{#2}#6}
\crefmultiformat{figure}{Figs.~#2#1\xdef\mycreffirstarg{#1}#3}{ and~#2({\crefstripprefix{\mycreffirstarg}{#1}}#3}{, #2({\crefstripprefix{\mycreffirstarg}{#1}}#3}{ and~#2({\crefstripprefix{\mycreffirstarg}{#1}}#3}

\newcommand{\blue}[1]{{\color{blue} #1}}

\newcommand{\Fig}[2][]{\cref{fig:#2}\if #1\empty\else(#1)\fi}

\newcommand\LDst{\bgroup\markoverwith{\blue{\rule[0.5ex]{2pt}{0.4pt}}}\ULon}

\begin{document}

\title{Shot-to-shot displacement noise in state-expansion protocols with inverted potentials}

\author{Giuseppe Paolo Seta }
\author{Louisiane Devaud}
\author{Lorenzo Dania}
\author{Lukas Novotny}
\author{Martin Frimmer}

\affiliation{Photonics Laboratory, ETH Z\"urich, 8093 Z\"urich, Switzerland}
\affiliation{Quantum Center, ETH Z\"urich, 8093 Z\"urich, Switzerland}

\date{\today}

\begin{abstract}
Optically levitated nanoparticles are promising candidates for the generation of macroscopic quantum states of mechanical motion. 
Protocols to generate such states expose the particle to a succession of different potentials. 
Limited reproducibility of the alignment of these potentials across experimental realizations introduces additional noise. 
Here, we experimentally investigate and model how such shot-to-shot noise limits the coherence length of a levitated nanoparticle during a state-expansion protocol using a dark, inverted electrical potential. 
We identify electric stray fields and mechanical instabilities as major sources of shot-to-shot fluctuations. 
We discuss the resulting experimental requirements for state expansion protocols exploiting inverted potentials. 

\end{abstract}

\maketitle

\paragraph{Introduction.—}
\label{sec:introduction}
The superposition principle is at the heart of quantum mechanics. It gives rise to interference effects that challenge our intuition, especially for massive objects. For example, electrons, atoms, and even macromolecules, have been delocalized over distances much larger than their own size~\cite{davisson1927scattering,Monroe1996schroedingerCat,Schumm2005matter-wave,hornberger2012quantum,fein2019quantum}. 
In a complementary approach, a mechanical mode of a comparatively massive crystal ($\sim10^{17}$~atoms) has been observed in a superposition state, albeit of extremely small size ($\sim10^{-18}~$m)~\cite{Bild2023schroedingerCat}.
Creating large superposition states of increasingly more massive systems is a prime goal of contemporary physics to test the limitations of quantum theory~\cite{bassi2013models}.

Optically levitated nanoparticles have been identified as a promising platform for this endeavor~\cite{romero2011quantumsuperposition,Romero-Isart2011largeQuantum_PRL,Bateman2014near-fieldNatComm,Wan2016PRLfreeNano,Nimmrichter2025electron-enabled,Bykov2025nanoparticlePRL}.
Mapping levitated systems onto the framework of optomechanics~\cite{Romero-Isart2011opticallyLevitating,Chang2010cavityoptomechanics,Aspelmeyer2014cavityRMP}, and understanding how to best detect their motion~\cite{tebbenjohanns2019optimal,Tebbenjohanns2020sideband}, 
have enabled cooling levitated oscillators to their motional ground state using optical cavities~\cite{delic2020cooling,piotrowski2023simultaneous,Ranfagni2022twodimensional}, and measurement-based feedback~\cite{magrini2021realtime,tebbenjohanns2021quantum,Kamba2022optical}.

The extent of the ground-state wavefunction of a nanoparticle (diameter \SI{100}{\nano\meter}) in a typical optical trap is around $\SI{10}{\pico\meter}$~\cite{magrini2021realtime,tebbenjohanns2021quantum}.
This small scale is a challenge for the generation of non-Gaussian states of motion~\cite{romero2011quantumsuperposition,Romero-Isart2011largeQuantum_PRL}. Such states require an anharmonic potential, which is currently not available on the minuscule length scale set by the zero-point fluctuations~\cite{gonzalez2021levitodynamics}.
Protocols exploiting anharmonicities on larger length scales~\cite{rodallordes2024macroscopic,romero2017coherent,neumeier2024fast} envision inflating the wavefunction to a size where the non-parabolic nature of the potential is sizable~\cite{romero2017coherent,Weiss2021largequantum,neumeier2024fast,Cosco2021enhancedPRA}. 

Importantly, achieving a large spatial position uncertainty $\sigma_z$ alone is not sufficient. Instead, preserving the quantum purity $\mathcal{P}$ of the state throughout the expansion is essential, making the coherence length $\xi = \sqrt{8}\mathcal{P}\sigma_z$ the relevant figure of merit~\cite{romero2011quantumsuperposition,romero2017coherent}.
The dominant source of decoherence in an optical trap in vacuum is photon recoil~\cite{Jain2016direct}, which inspired proposals harnessing ``dark'' potentials that rely on electric or magnetic fields~\cite{rodallordes2024macroscopic,Pino2018on-chipQuantum}.

On the experimental front, a number of efforts have been taken towards state expansion.
All experiments initialize the particle in a tight optical trap, which is then either switched to a weaker confining potential~\cite{bonvin2024state,rossi2024quantum}, or switched off for free-evolution~\cite{hebestreit2018sensing,mattana2025traptotrapfreefallsoptically,steiner2025freeexpansionchargednanoparticle,Kamba2023revealing,Kamba2025quantum}, or switched to an inverted configuration for exponentially accelerated expansion~\cite{tomassi2025accelerated,duchavn2025nanomechanical}. 
Efforts entering the quantum regime thus far rely on purely optical potentials~\cite{rossi2024quantum,Kamba2025quantum}.
Dark potentials, which promise expansion beyond the photon-recoil limit, have been used to expand thermal states of levitated nanoparticles  in electrical confining and inverted configurations~\cite{bonvin2024state,tomassi2025accelerated}. 
This progress has been enabled by hybrid traps, combining optical levitation with ion-trapping technology~\cite{conangla2018motion,conangla2020extending,bykov2022hybrid,bonvin2024hybrid}.

Crucially, all protocols that expose the particle to different potentials require their repeatable alignment with sufficient precision~\cite{gonzalez2021levitodynamics}. 
Several proposals have recognized that a variation of this alignment (i.e., displacement noise) from one execution of the experiment to the next may limit demonstrations of quantum interference~\cite{Pedernales2022robustPRA,rodallordes2024macroscopic,neumeier2024fast}, and first evidence of this effect has recently surfaced in experiments with optical potentials~\cite{rossi2024quantum}.
In that light, it is surprising that no experimental study has systematically scrutinized displacement noise in levitation experiments, especially in configurations that deploy both bright and dark potentials.

In this work, we investigate shot-to-shot noise in a protocol expanding the motional state of an optically levitated nanoparticle in a dark inverted electrical potential.
We find that mechanical instabilities and electrical stray forces limit the particle's coherence length. 
Our model matches our experimental observations and reveals that shot-to-shot noise is different in nature as compared to white-noise. 
For currently available experimental parameters (initial coherence length 20~pm), reaching a coherence length of 1~nm requires a shot-to-shot displacement stability better than 0.5~pm.
We discuss the challenges associated with shot-to-shot displacement noise in the context of levitodynamics, along with possible mitigation strategies.

%%% Fig. 1  %%%%%%%%%%%%%%%%%%%%%%%%%%%%%%%%%%%%%%%%
\begin{figure*}[ht]
    \includegraphics[width=1\textwidth]{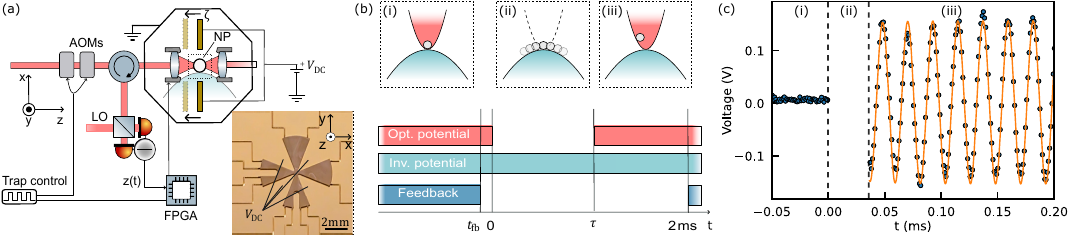}
    \caption{(a)~Experimental setup. A nanoparticle (NP) is trapped by a focused laser beam in vacuum. The laser is toggled by two acousto-optic modulators (AOMs). The particle's motion is detected by mixing back-scattered light with a homodyne local oscillator (LO). The position signal is processed by a field-programmable gate array (FPGA) and used for feedback cooling.
    A planar chip with electrodes (see inset) is placed in the focal plane of the optical trap. The chip position $\zeta$ can be translated with a piezo stage (not shown). The electrodes are connected to a common static voltage $V_{\mathrm{DC}}$, generating an inverted potential along the $z$ axis.
    (b)~State expansion protocol.
    Phase (i): Initialization in the optical potential, which dominates over the electrical potential, which is always on.
    Cooling is turned off at $t_\mathrm{fb}=-\SI{1}{\micro \second}$ before the optical trap is disabled at $t=0$. 
    Phase (ii): evolution in the inverted potential for a time $\tau$.
    Phase (iii): recapture and measurement in the optical trap.
    (c)~Detector time trace for a single realization of the protocol, where the particle evolves in the inverted potential for $\tau=\SI{40}{\micro\second}$.
    The raw trace is filtered (orange line) to estimate the recapture position  $z(\tau)$ and momentum $p_{\rm z}(\tau)$.}
    \label{fig:setup}
\end{figure*}
%%%%%%%%%%%%%%%%%%%%%%%%%%%%%%%%%%%%%%%%%%%%%%%%%%%%

\paragraph{Experimental details.---}
Figure~\ref{fig:setup}(a) shows a sketch of the experimental setup, which consists of an optical trap and an electrical inverted potential.
We trap single charged silica nanoparticles of nominal radius $R = \SI{78}{\nano\meter}$, charge $|q|\approx 100e$, and mass $m\approx \SI{4.4}{\femto\gram}$ at pressure $p \approx \SI{7e-7}{\milli\bar}$. 
The optical trap is generated by a laser (wavelength $\lambda = \SI{1550}{\nano \meter}$, power $P \approx \SI{600}{\milli \watt}$, optical axis along $z$, polarization along $x$, gravity along $y$) focused by an aspherical lens (numerical aperture NA = 0.8).
The optical gradient force generates a three-dimensional harmonic potential for the c.m.\ motion of the trapped particle with trap frequencies $\left(\Omega^z_o,\Omega^x_o,\Omega^y_o \right) = 2\pi \times  (44, 131, 150)~\mathrm{kHz}$.

We detect the light backscattered by the particle \cite{tebbenjohanns2019optimal} to apply linear feedback cooling to all three c.m.\ degrees of freedom to reduce their thermal population~\cite{tebbenjohanns2019cold}. Throughout this work, we focus on the $z$ mode.
The lenses that focus and recollimate the trapping laser are mounted in conductive holders, which double as electrodes to apply a force along $ z$ to the charged nanoparticle.
One of these electrodes applies an AC voltage for feedback cooling along the $z$ axis, while the other is set to a constant potential to compensate for stray fields. 
Two acousto-optic modulators (AOMs) are used to turn the optical trap off and on, with a switching time of \SI{200}{\nano\second}. 

The electric potential is generated by a micro-machined chip~\cite{bonvin2024hybrid}. 
Four blade electrodes are carved out of a planar substrate (thickness $\SI{300}{\micro\meter}$) and metalized. The separation between the tips of the electrodes and the nanoparticle is $\SI{250}{\micro \meter}$. We apply a common static voltage $V_\text{DC}$ to all four blade electrodes. Choosing the polarity of this voltage to match the polarity of the particle's net charge produces an anti-confining potential along the $z$ axis and a confining potential in the $xy$ plane.
To leading order in $z$, the electric potential along the optical axis is an inverted harmonic potential U=$-m\Omega_{\rm inv}^2z^2$.
The frequency $\Omega_{\rm inv}$ characterizes the inverted potential curvature and is set by the voltage $V_{\rm DC}$.  
The chip is mounted on a three-axis linear piezo stage to align the inverted electrical potential relative to the optical trap.

We expand the motional state of the levitated particle along the $z$ axis by evolution in the inverted electrical potential. 
To this end, we keep the electrical potential continuously on during the experiment, while the optical trap is switched off and on. When the optical trap is on, the anti-confining electrical potential along $z$ superposes with the confining optical potential. The frequencies of the inverted electrical potential are always below $\Omega_\text{inv}/(2\pi) = \SI{13}{\kilo\hertz}$, and therefore sufficiently small to lead to a net confining potential in presence of both the optical and the electrical potential.

\paragraph{Expansion protocol.---}

The experimental sequence for state expansion is composed of three phases, illustrated in Fig.~\ref{fig:setup}(b).
In phase (i), both the optical trap and the inverted electrical potential are engaged, leading to a net confining potential with frequency $\Omega_o^z$. The particle is initialized by feedback cooling in the optical trap to a state with position variance $\sigma_0^2$. 
In phase (ii), the optical trap is turned off and the particle evolves in the inverted electrical potential for a time $\tau$.  
In phase (iii), the optical trap is switched back on and the particle is recaptured in the confining potential in absence of feedback.

In Fig.~\ref{fig:setup}(c), we plot an example of a measured position time trace of the particle along $z$ for an evolution in the inverted potential for $\tau=\SI{40}{\micro\second}$.
The position data during phase (i) represent the particle trajectory under feedback cooling. During phase (ii), the light field is off and no measurement record is available. Position data during phase (iii) hold information about the motional state of the particle after evolution in the inverted potential.
We apply a retrodiction filter for times $t>\tau$ to optimally estimate the particle trajectory \cite{rossi2019observing,bonvin2024state}.
From the filtered trace, we extract the recapture position $z(\tau)$ and momentum $p_z(\tau)$ in the optical trap at recapture time $t=\tau$. 
The protocol is repeated 200 times to acquire the mean position $\langle z\rangle$, momentum $\langle p_z\rangle$, their respective standard deviations $\sigma_z$ and $\sigma_{p_z}$, as well as their covariance $\sigma_{z,p_z}^2$ at time $\tau$.

%%% Fig. 2  %%%%%%%%%%%%%%%%%%%%%%%%%%%%%%%%%%%%%%%%
\begin{figure}[tb]
    \centering
    \includegraphics[width=1\columnwidth]{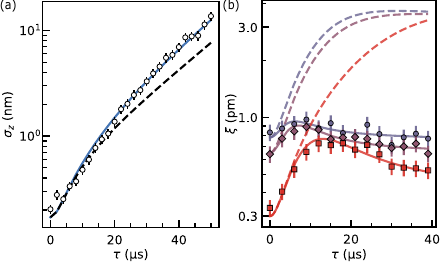}
    \caption{    
    (a)~Measured standard deviation of recapture positions $\sigma_z$ as a function of expansion time $\tau$.
    The black dashed line is the model restricted to Eqs.~\eqref{eq:coh_variance} and~\eqref{eq:heat_variance}, i.e., in absence of shot-to-shot noise.
    The blue continuous line is a fit to our full model that includes shot-to-shot noise according to Eq.~\eqref{eq:shottoshot}. 
    (b)~Inferred coherence length $\xi(\tau)$ for three different initial state sizes $\sigma_0 =\SI{150}{\pico\meter}$ (blue circles), $\sigma_0 =\SI{187}{\pico\meter}$ (purple diamonds), and $\sigma_0 =\SI{384}{\pico\meter}$ (red squares). The solid lines are fits to our full model including shot-to-shot noise. The dashed lines show the model prediction neglecting shot-to-shot noise. 
    }
    \label{fig:expansion}
\end{figure}
%%%%%%%%%%%%%%%%%%%%%%%%%%%%%%%%%%%%%%%%%%%%%%%%%%%

To illustrate the expansion, Fig.~\ref{fig:expansion}(a) shows the measured standard deviation $\sigma_z$ as a function of evolution time $\tau$ in an inverted potential with frequency $\Omega_{\rm inv}=2\pi\times\SI{9.3}{\kilo\hertz}$.
The value of $\sigma_z$ expands from $\sigma_z(0)=\SI{202 \pm 21}{\pico\meter}$ ($\approx 450$~phonons) up to $\sigma_z(\SI{50}{\micro\second})=\SI{13.7 \pm 1.42}{\nano\meter}$, corresponding to an increase by $\SI{37}{\decibel}$ in variance.

%%%%%%%%%%%%%%%%%%%%%%%%%%%%
% insert coherence length
%%%%%%%%%%%%%%%%%%%%%%%%%%%%%

%\paragraph{Expansion of coherence length.—}
For future experiments in the quantum regime, the relevant quantity is not the total position uncertainty $\sigma_z$, but rather its coherent part, which is quantified by the coherence length $\xi = \sqrt{8}\mathcal{P}\sigma_z$~\cite{romero2011quantumsuperposition}.
For a Gaussian state, the purity is given by $\mathcal{P} =\hbar/[2(\sigma^2_z\sigma^2_{p_z} - \sigma^4_{zp_z})^{1/2}]$. It measures how far the state is from the Heisenberg bound. 
In Fig.~\ref{fig:expansion}(b), we show as blue circles the coherence length $\xi$ inferred from our measurements as a function of $\tau$ for an inverted potential with frequency $\Omega_{\mathrm{inv}} = 2\pi \times \SI{7.6}{\kilo\hertz}$ starting from an initial state size $\sigma_0=\SI{150}{\pico\meter}$ (corresponding to an initial coherence length $\xi_0=\SI{0.83}{\pico\meter}$). 
The inferred coherence length slightly increases with increasing $\tau$, reaching a maximum of $\SI{0.97}{\pico\meter}$, before decaying. 
As a second example (red squares), we show the same experiment with a larger initial state size $\sigma_0=\SI{384}{\pico\meter}$ (corresponding to $\xi_0=\SI{0.33}{\pico\meter}$). This larger state size is obtained by reducing the gain of the feedback cooling. 
The inferred coherence length of this comparably hot state expands to $\xi=\SI{0.73}{\pico\meter}$ in the inverted potential before decaying. 
We show a third dataset with an intermediate starting state size $\sigma_0=\SI{187}{\pico\meter}$ [purple diamonds]. 
%

%%%%%%%%%%%%%%%%%%%%%%
% end insert coh length
%%%%%%%%%%%%%%%%%%%%%%%%%%%%

\paragraph{White noise heating.---}
We proceed by comparing our measurements to the currently established model for the state size $\sigma_z$ and coherence length $\xi$~\cite{bonvin2024state, tomassi2025accelerated}.
The total variance $\sigma_z^2(\tau)=\sigma^2_{\rm coh}(\tau)+\sigma^2_{\rm wn}(\tau)$ is the sum of two terms. The coherent part 
\begin{equation}
\sigma_{\rm coh}^2(\tau) = \sigma_0^2 \left[\cosh^2(\Omega_{\rm inv}\tau) + r^2 \sinh^2(\Omega_{\rm inv}\tau)\right]
\label{eq:coh_variance}
\end{equation}
represents the desired expansion of the initial variance $\sigma_0^2$ in the inverted potential, where $r=\Omega_o^z/\Omega_\text{inv}$ is the ratio of the confining and inverted potential frequencies. 
The incoherent part 
\begin{equation}
\sigma_{\rm wn}^2(\tau) = \frac{r\hbar\Gamma}{m\Omega_{\rm inv}}\left[\frac{\sinh(2\Omega_{\rm inv}\tau)}{2\Omega_{\rm inv}}-\tau   \right]
\label{eq:heat_variance}
\end{equation}
represents added position variance due to white-noise heating by a thermal bath at a rate $\Gamma$ (measured in units of phonons of the optical potential). In practice, $\Gamma$ stems from photon recoil and gas collisions.
For our system, from an independent reheating measurement~\cite{tebbenjohanns2019cold}, we extract $\Gamma=2\pi \times\SI{554}{\kilo\hertz}$, matching the value expected from collisions with gas molecules at the measured pressure.

To compare theory and experiment, we plot (black dashed line) in Fig.~\ref{fig:expansion}(a) the calculated standard deviation $\sigma_z$ from Eqs.~\eqref{eq:coh_variance} and~\eqref{eq:heat_variance} using the measured value for $\Gamma$ and $\sigma_0=\SI{170}{\pico\meter}$.
The measured position standard deviation exceeds the one predicted by coherent expansion and white-noise heating. 
Even more striking is the result when adding the model prediction for the coherence length to Fig.~\ref{fig:expansion}(b).
The model (dashed lines) clearly overestimates the experimentally inferred coherence length at long times, suggesting an additional source of noise not accounted for by the model.

A particularly revealing feature in Fig.~\ref{fig:expansion}(b) is that the experimentally inferred coherence lengths at large $\tau$ approach different values for different initial coherence lengths $\xi_0$. 
In contrast, state expansion in an inverted potential under white-noise heating drives the coherence length to a value that is independent of the initial coherence length. 
This can be observed in Fig.~\ref{fig:expansion}(b), where the dashed lines asymptotically approach the same value at large $\tau$. 
Therefore, our results speak against the validity of any model that considers only white-noise heating.

In the following, we provide evidence that the observed additional noise is due to shot-to-shot fluctuations of the relative alignment of the optical and the effective electrical potential.

\paragraph{Shot-to-shot displacement noise.---}
When the relative displacement between the inverted potential and the confining potential is constant during a single realization of the expansion protocol, but varies between realizations, the additional shot-to-shot position variance
\begin{equation}
\sigma^2_{\rm shot} =\sigma^2_{\rm disp} \left(r^{-2}+1\right)^2\left[1-\cosh(\Omega_{\rm inv}\tau)\right]^2
\label{eq:shottoshot}
\end{equation}
contributes to the state size, where $\sigma^2_{\rm disp}$ is the shot-to-shot variance of the spatial displacement between optical and inverted potential (see Supplement for derivation~\cite{supplement}).

%%%%%%%%%%%%
% insert discussion of ss noise model
%%%%%%%%%%%%%%%%

We return to the data in Fig.~\ref{fig:expansion}(b) and fit each dataset to our full model with $\sigma_0$ and $\sigma_{\mathrm{disp}}$ as the only free parameters (solid lines). For $\sigma_{\rm disp}$, we obtain the values $3.68\pm0.53$~nm (circles), $3.38\pm0.42$~nm (diamonds), and $2.27\pm0.12$~nm (squares). 
The agreement between the measured data and the fit strongly supports our hypothesis that shot-to-shot noise makes a significant contribution in our experiments.
Also the measured position standard deviation in Fig.~\ref{fig:expansion}(a) matches our full model (blue solid line) with the fit parameters $\sigma_{\rm disp}  = \SI{1.14 \pm 0.07}{\nano\meter}$ and $\sigma_0=\SI{170 \pm 8}{\pico\meter}$.

%%%%%%%%
% end insert
%%%%%%%%%%%%%%%%%%

We continue by identifying the different contributions to the shot-to-shot displacement noise $\sigma_\text{disp}$. 
First, an electric stray field acts as a Coulomb force on the particle. 
This stray force effectively shifts the optical and the electrical potential relative to each other. 
We denote the shot-to-shot variance of this electrical stray force by $\sigma^2_{\rm sf}$.  Second, the position of the electrical potential, set by the position of the chip relative to the optical potential (quantified by their distance $\zeta$), can fluctuate, which is quantified by the shot-to-shot chip-position variance $\sigma_{\zeta}^2$. 
Assuming uncorrelated chip-position and stray-force fluctuations, the effective displacement variance is
\begin{equation}
\sigma^2_{\rm disp}=\sigma^2_{\rm sf}(m\Omega^2_{\rm inv})^{-2}+ \sigma^2_{\zeta}.
\label{eq:sigmadisp}
\end{equation}
In the first term, the force fluctuations are transduced into position fluctuations by the spring constant of the inverted potential.

\begin{figure}[t]
    \includegraphics[width=1\columnwidth]{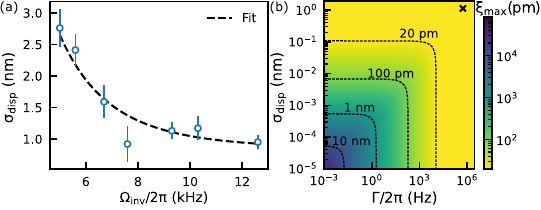}
    %{figures/Fig3_dirtyMF.pdf}
    \caption{
    (a)~Measured $\sigma_{\rm disp}$ as function of $\Omega_{\mathrm{inv}}$. Error bars indicate uncertainty of the fit used for extraction. The dashed line is a fit of the data to Eq.~\eqref{eq:sigmadisp}.
    (b)~False-color plot of calculated maximally achievable coherence length $\xi_\text{max}$ as a function of $\sigma_{\mathrm{disp}}$ and $\Gamma$. We fix $\sigma_0=\SI{7}{\pico\meter}$, $\Omega_o = 2\pi\times\SI{40}{\kilo\hertz}$ and $\Omega_{\rm inv} = 2\pi\times\SI{10}{\kilo\hertz}$. 
    Dotted lines are contours for the indicated coherence lengths. 
    The black cross represents the values of $\sigma_{\rm disp}$ and $\Gamma$ for the measurement in Fig.~\ref{fig:expansion}(a). }
    \label{fig:coherence_measurement}
\end{figure}

To individually quantify $\sigma_{\rm sf} $ and $\sigma_{\zeta}$, we repeat the experiment in Fig.~\ref{fig:expansion}(a) for different values of the inverted potential frequency $\Omega_{\rm inv}$. 
We plot the obtained values of $\sigma_{\mathrm{disp}}$ as a function of $\Omega_{\rm inv}$ in Fig.~\ref{fig:coherence_measurement}(a) and observe that $\sigma_{\mathrm{disp}}$ drops with increasing $\Omega_\text{inv}$.
To quantify the different shot-to-shot noise contributions, we fit (dashed line) the data to Eq.~\eqref{eq:sigmadisp}, and extract the parameters $\sigma_\mathrm{sf} = \SI{10.9 \pm 0.5}{\atto\newton}$ and $\sigma_{\zeta} = \SI{834 \pm 53}{\pico\meter}$. 
We conclude that $\sigma_\text{disp}$ is dominated by stray force fluctuations for small $\Omega_{\rm inv}$, whereas mechanical position fluctuations prevail for large $\Omega_{\rm inv}$.

%%% Fig. 3  %%%%%%%%%%%%%%%%%%%%%%%%%%%%%%%%%%%%%%%%
           
 %%%%%%%%%%%%%%%%%%%%%%%%%%%%%%%%%%%%%%%%%%%%%%%%%%%

\paragraph{Discussion.---}
We now use our model to assess the noise conditions required for expanding the motional state of a levitated nanoparticle to dimensions comparable to the particle size.
In Fig.~\ref{fig:coherence_measurement}(b), we show a false-color plot of the maximally achievable coherence length $\xi_\text{max}$ for an expansion of a particle in its ground-state (zero-point size $\sigma_0= \SI{7}{\pico\meter}$, corresponding optical trap frequency $\Omega_o=2\pi\times\SI{40}{\kilo\hertz}$) in an inverted potential with frequency $\Omega_{\mathrm{inv}} = 2\pi \times \SI{10}{\kilo\hertz}$ as a function of $\sigma_{\rm disp}$ and $\Gamma$.
We observe that reaching $\xi_\text{max} = \SI{1}{\nano\meter}$ requires $\Gamma/(2\pi) < \SI{1}{\hertz}$ and $\sigma_{\mathrm{disp}} < \SI{0.5}{\pico\meter}$. 
Therefore, expanding the coherence length to $\xi\sim50 \xi_0$ requires a shot-to-shot displacement noise that is $\sim 50$ times smaller than the initial coherence length ($\sigma_\text{disp} \lesssim \xi_0/50$).
This value is comparable to the stability requirement obtained by analyzing the visibility of interference fringes in an all-optical interference protocol~\cite{neumeier2024fast}.
Regarding the scaling, an enhancement of $\xi_\text{max}$ by one order of magnitude requires an improvement of $\sigma_{\mathrm{disp}}$ by roughly one order of magnitude and $\Gamma$ by two orders of magnitude.

We can now deduce technical requirements. 
For white-noise heating by electric fields, the voltage-noise spectral density required for a desired $\Gamma$, at a given electrode distance $d$, can be estimated as $S_{\rm v} = 2d^2\Gamma \hbar m \Omega_o/q^2$~\cite{brownnutt2015ion}. To achieve $\Gamma/(2\pi) = \SI{1}{\hertz}$ with our particle and assuming $d = \SI{1}{\milli\meter}$, we obtain $S_{\rm v} \approx \SI{6}{\nano\volt\per\sqrt\hertz}$. This value is comparable to $\SI{10}{\nano\volt\per\sqrt\hertz}$, achievable in practice~\cite{dania2024ultrahigh,vinante2019testing}.

Let us turn to the shot-to-shot noise $\sigma_{\mathrm{disp}}$ and its two contributions $\sigma_\zeta$ and $\sigma_\text{sf}$. 
For the mechanical shot-to-shot fluctuations $\sigma_\zeta$ to remain below 0.5~pm in a bandwidth set by a realistically required duration of the expansion of $\tau_\text{ex}=\SI{100}{\micro\second}$ requires a position noise level $S_{\zeta\zeta} = 2\tau_{\mathrm{ex}}(\SI{0.5}{pm})^2\approx \SI{e-28}{\meter\squared\per\hertz}$ in the low frequency range.
Encouraging experiments with superconductive traps and levitated magnetic particles have demonstrated mechanical noise down to $\SI{e-27}{\meter\squared\per\hertz}$ in the few-Hertz range~\cite{fuchs2024measuring}.
Following the same reasoning, low-frequency stray forces have to be suppressed to a noise level of $S_{\rm sf}=2\tau_{\rm ex}(m\Omega^2_{\rm inv}\times\SI{0.5}{\pico\meter})^2  \approx \SI{e-44}{\newton\squared\per\hertz}$ for $\Omega_{\rm inv}=2\pi\times\SI{10}{\kilo\hertz}$. 
To achieve the required $S_{\rm sf}$, it is crucial to understand the stray force origin. 
We suspect that the main contribution to $\sigma_\text{sf}$ stems from charge fluctuations of dielectric surfaces in close proximity to the particle, such as the trapping and collection lenses~\cite{brownnutt2015ion,teller2021heating,sagesser20243dimensional,hebestreit2018sensing}. These fluctuations may be reduced by coating the lens surfaces with conductive transparent materials.

Besides passively reducing stray fields and mechanical instabilities, one can envision measuring their instantaneous values at each shot of the experiment to either correct for them actively or in post-processing.
Furthermore, dedicated protocols to cancel shot-to-shot noise in expansion protocols developed for confining traps may find adaptations for inverted potentials~\cite{rossi2024quantum}.

Finally, we comment on the choice of inverted trap frequency, which is not straightforward in the parameter range of interest (see Supplement for details~\cite{supplement}). 
In the absence of shot-to-shot fluctuations, where noise is purely due to white-noise heating, it is always beneficial to maximally accelerate the expansion dynamics by choosing $\Omega_\text{inv}$ large. 
In contrast, in the presence of shot-to-shot noise, larger $\Omega_\text{inv}$ leads to a reduction in maximally achievable coherence length for constant $\Gamma$ and $\sigma_\text{inv}$. Thus, for displacement noise from electrode position fluctuations [$\sigma_\zeta$ in Eq.~\eqref{eq:sigmadisp}], higher $\Omega_\text{inv}$ is undesirable. 
However, the contribution of stray forces in Eq.~\eqref{eq:sigmadisp} is reduced by the inverse trap frequency $\Omega_\text{inv}^4$. Thus, when limited by stray forces, increasing $\Omega_\text{inv}$ is beneficial. 
Therefore, a quantitative characterization of the different sources of noise is required to optimally choose experimental parameters.

\paragraph{Conclusions.---}
We have experimentally quantified the sources of noise acting on a levitated nanoparticle during a state-expansion protocol in a dark inverted potential. 
%We have analyzed the sources of decoherence in a dark inverted electrical potential.
Besides white-noise heating, we have identified shot-to-shot displacement fluctuations between experimental repetitions as the dominant limitation of the coherence length.
These shot-to-shot fluctuations have contributions from mechanical motion of the chip holding the electrodes, and from fluctuations of electric stray fields.
Our model for shot-to-shot noise is supported by our experimental findings. 
With this model, we have quantified the required shot-to-shot stability for reaching coherence lengths in the nanometer range. 
The relevance of our work extends across the diverse communities harnessing levitated nanoparticles in vacuum. 
It provides a starting point for the experimental progress that will be required to enable matter-wave experiments using optical and electrical potentials~\cite{romero2011quantumsuperposition,Romero-Isart2011largeQuantum_PRL,rodallordes2024macroscopic,neumeier2024fast,Bateman2014near-fieldNatComm}. 
This progress will also benefit the communities aiming to exploit squeezed states of levitated motion for quantum sensing~\cite{Weiss2021largequantum,Cosco2021enhancedPRA,rossi2024quantum,Kamba2025quantum}, as well as endeavors aiming to couple levitated nanoparticles to other quantum systems, such as free electrons~\cite{Nimmrichter2025electron-enabled} or two-level systems~\cite{Wan2016PRLfreeNano,Bykov2025nanoparticlePRL}.
Finally, this work may serve as a platform to interface with the quantum-simulation community, currently facing shot-to-shot noise as a serious technical concern~\cite{Steckmann2025error}.

\paragraph{Acknowledgments.---}
This research has been supported by the Swiss SERI Quantum Initiative (grants no. UeM019-2 and no. UeM029-3), by the Swiss National Science Foundation (grant no. 212599), and by the European Research Council (ERC) under grant agreement no. 951234 (Q-Xtreme ERC 2020-SyG). L. Dania acknowledges support from the Quantum Center Research Fellowship and the Dr Alfred and Flora Spälti Fonds. L. Devaud thanks for support through her SNSF Fellowship (TMPFP2\_217122). We thank the participants of the Q-Xtreme retreats as well as C. Dellago for fruitful discussions.

\bibliography{biblio_noises_contributions}

\end{document}